\documentclass{article}[12 pt]
\newcommand{\be}{\begin{equation}}
\newcommand{\ee}{\end{equation}}
\newcommand{\bea}{\begin{eqnarray}}
\newcommand{\eea}{\end{eqnarray}}
\begin{document}


\renewcommand{\thefootnote}{\fnsymbol{footnote}}
\renewcommand{\baselinestretch}{1.3}
\medskip

\begin{center}
{\large {\bf 2-D Midisuperspace Models for Quantum Black Holes }}
\\ \medskip  {}
\medskip

\renewcommand{\baselinestretch}{1}
{\bf
J. Gegenberg
\\}
\vspace*{0.50cm}
{\sl
 Dept. of Mathematics and Statistics and Department of Physics,
University of New Brunswick\\
Fredericton, New Brunswick, Canada  E3B 5A3\\ [5pt]
}
{\bf G. Kunstatter
\\}
\vspace*{0.50cm}
{\sl
 Department of Physics,
University of Winnipeg\\
Winnipeg,Manitoba, Canada R3B 2E9\\ [5pt]
}
\end{center}

\renewcommand{\baselinestretch}{1}

\begin{center}
{\bf Abstract}
\end{center}
Dimensionally reduced spherically symmetric gravity and its generalization, generic 2-D dilaton gravity, provide ideal theoretical laboratories for the study of black hole quantum mechanics and thermodynamics. They are sufficiently simple to be tractable but contain enough structure to allow the study of many deep issues in quantum gravity, such as the endpoint of Hawking radiation  and the source of black hole entropy. This article reviews recent progress in a particular geometrical approach to the study of quantum black holes in generic 2-d dilaton gravity.
{\small
}\vfill
\hfill January 2009 \\

\newpage

\section{Introduction}\label{sec1.1}

Despite significant progress in recent years, particularly in the context of string theory\cite{strings} and loop quantum gravity\cite{thiemann_book}, the nature of quantum gravity remains the great unsolved mystery of modern theoretical physics. With the possible exception of some brane-world scenarios, most conceivable versions of a quantum theory of gravity will likely only be relevant at energy and distance scales that will remain experimentally inaccessible in the forseeable future. Most research in the field consists of an examination of the internal self-consistency of given candidate theories as well as their potential for solving the outstanding theoretical problems in classical and semi-classical general relativity. These theoretical problems include resolution of the singularities predicted by general relativity as well as the black hole entropy and information loss puzzles.

In order to make any theory of quantum gravity amenable to rigorous quantitative analysis, it is necessary to simplify the equations by focussing on a sector of the theory that is thought to contain enough structure to access the relevant issues. In this regard, the most promising arenas of study are cosmology, in which homegeneity and isotropy are good approximate symmetries, and black holes for which the no-hair theorem suggests that spherical symmetry provides a reasonable starting point\footnote{Strictly speaking, one should consider axial symmetry in this context, but for in many cases, zero angular momentum should be a good approximation.}. This review will deal only with black holes, whose thermodynamic properties suggest that they contain valuable clues about the underlying microscopic theory of quantum gravity.

One particular model that has been extensively studied in the context of quantum black holes is  dimensionally reduced spherically symmetric gravity\cite{berger73,unruh76}. The Birkhoff theorem guarantees that the vacuum theory is dynamically trivial (there are no propagating spherically symmetric  graviton modes) but the model is diffeormorphism invariant and hence maintains the essential kinematical features of the full theory, therebye providing a valuable ``midi-superspace model'' for quantization.

The class of models that we will describe generalizes spherically symmetric gravity to a large class of diffeomorphism invariant theories in two space-time dimensions. These theories all satisfy a Birkhoff theorem and allow for the existence of black hole solutions with the same thermodynamic properties (entropy, semi-classical temperature) as physical black holes in higher dimensions. They are collectively known as generic 2-D dilaton gravity.

Few researchers have made more substantial and important contributions to the study of generic 2-D dilaton gravity than Professor Kummer and his collaborators. An excellent description of these contributions is contained in the very thorough review written by D. Grumiller, W. Kummer and D.V. Vassilevich\cite{grumiller02}. The present article cannot hope to compete with the completeness and depth of that review. Instead, we will concentrate on the small part of the field with which we are most familiar. In particular, we will focus on analyses in terms of the geometrodynamical variables of the theory: the metric and the dilaton field. In the context of spherically symmetric gravity, the dilaton has a geometrical interpretation as the area of the Killing sphere at fixed radius from the origin. If one is considering a theory that cannot be derived from higher dimensional gravity, the dilaton still plays a crucial role in determining the thermodynamic properties of black holes in the generic theory. This will be described in more detail below.

The paper is organized as follows: In the next section we will provide a brief introduction to the class of models that we are considering, including the action and general solution. Section III will describe how the physical observables and thermodynamic properties can be extracted in a very simple, diffeomorphism invariant manner. Section IV will present the Hamiltonian analysis, including the complete reduction to the physical phase space, which will be a precursor to the study in Section V of its semi-classical quantum properties. Section VI will describe the Dirac quantization of the theory, showing that the Hamiltonian constraint can be solved to reveal interesting quantum structure. Finally, we close with conclusions and prospects for future work, particularly the possibility of incorporating matter so as to provide a self-consistent treatment of the quantum dynamics of gravitational collapse in the generic theory.

\section{Generic 2-D Dilaton Gravity: Action and Solutions}

 The gravitational action we wish to consider is:
\be
S_G[g,\phi]=\frac{1}{2G}\int dx dt \sqrt{-g}\left(\phi R(g)+{V(\phi)\over l^2}\right),\label{dgaction}
\ee
where $l$ is a parameter with dimensions of length which is generally taken to be the Planck length in the theory. Note that the generic theory is completely specified by the form of the dilaton potential $V(\phi)$. This action is the most general diffeomorphism invariant action in two space-time dimensions that has at most second derivatives of the fields\cite{banks91}. Note that there is no kinetic term for the scalar field in (\ref{dgaction}).  Had we chosen to add such a term, it could always be removed by a conformal reparametrization of the metric of the general form:
\be
g_{\mu\nu}\to \Omega(\phi)g_{\mu\nu}
\ee
Such reparametrizations leave the conformal structure of the geometry invariant, but do affect geodesics of test particles. In the cases of interest to be described below, the physical metric is  in fact related to $g_{\mu\nu}$ above by precisely such a conformal reparametrization, one that is regular everywhere except at the curvature singularity of the physical metric.

 As expected from spherically symmetric gravity, the generic vacuum theory has no propagating modes. There is a one parameter family of classical solutions with a single Killing vector\cite{louis-martinez94}. In Schwarzschild-like coordinates in which the dilaton is used as the spatial coordinate the solution takes the form:
\bea
\phi &=& x/l,\nonumber\\
ds^2 &=& -(j(\phi)-2GlM)dt^2 + (j(\phi)-2GlM)^{-1}d\phi^2,
\label{metric}
\eea
where
\be
j(\phi):=\int^\phi_0 d\tilde{\phi} V(\tilde{\phi}).
\ee
As will be discussed in the next section, the existence of black hole solutions in the theory and their corresponding conformal properties depend on the form of the dilaton potential $V(\phi)$ and its first integral $j(\phi)$.

Specific dilaton gravity theories with action equivalent (up to local reparametrizations) to $S_G$ were considered extensively in the past. One of the earliest was proposed in 1984 by Jackiw and Teitelboim\cite{jt}. The Jackiw-Teitelboim theory has dilaton potential $V(\phi) = \lambda\phi$.
 This theory came into further prominence when it was realized that it is equivalent to the cylindrically symmetric dimensionally reduced 2+1 gravity with cosmological constant $\lambda$, which also lacks local propagating modes but nevertheless has interesting black hole solutions \cite{BTZ}. The vacuum theory can be reduced via dimensional reduction to J-T coupled to an Abelian gauge field theory.

Another theory that received a great deal of attention in the early 1990's because of its connection to string theory is the 2-d vacuum dilatonic black hole (the so-called Witten black hole\cite{witten91} in the string motivated CGHS model\cite{cghs}. A thorough analysis of the thermodynamic properties of the Witten black holes can be found in Bose {\it et al}\cite{louko95}. As stated above, this theory is exactly solvable both classically and quantum mechanically, and there was hope that it would provide clues about the back reaction and end-point of the collapse/radiation process (see for example Mann\cite{mann93} and Bose {\it et al}\cite{bose95} and references therein.).

\section{Classical Vacuum Theory: Observables and Thermodynamics}

The beauty of 2-d dilaton gravity is that it is simple enough to be tackled generically. A systematic analysis of the generic theory was undertaken in the early to mid 1990's by several groups using a variety of techniques.  One can prove a Birkhoff theorem \cite{louis-martinez94} for arbitrary potential $V(\phi)$, i.e. for the most general theory, and explicitly write down all the solutions in terms of a single physical parameter, which can be interpreted as the total energy.  The physical observables and thermodynamics properties associated which such black hole solutions were initially derived for various specific 2-d theories\cite{mann93,lemos93}, but it is possible to do a completely general analysis for the generic action in Eq.(\ref{dgaction}) as well \cite{gegenberg94a, louis-martinez93}. The basis for such an analysis is existence of a global Killing vector that can be written down in covariant form for the generic theory in terms of the dilaton\cite{gegenberg94a}:
\be
k^\mu = \eta^{\mu\nu}\phi_{,\nu}.
\label{killing field}
\ee
The vanishing of the norm of this Killing vector signals as usual the presence of a Killing horizon. When the spatial slice $\phi=0$ is excluded from the spacetime this Killing horizon is indeed an event horizon that provides a boundary between the interior of the black hole and the asymptotic region. In the context of the metric $g_{\mu\nu}$ it is not obvious that the surface $\phi=0$ has a curvature singularity and indeed for some solutions this surface is completely regular.
In this regard, one observes that the metric (\ref{metric}) is not generically asymptotically flat, but if $j(\phi)$ diverges for large $\phi$, one can define a physical metric by a conformal rescaling:
\be
ds^2_{phys} = {1\over j(\phi)}ds^2= -(1-2GM l/j(\phi))dt^2 + (1-2GMl/j(\phi))^{-1}\left({d\phi\over j(\phi)}\right)^2.
\label{physical metric}
\ee
The physical metric (\ref{physical metric}) is asymptotically flat and has a singularity at $\phi=0$.
When the dilaton gravity theory corresponds to the spherical reduction of D-dimensional gravity, this physical metric is precisely the radial part of the higher dimensional metric, as can be verified by changing coordinates to $\phi = r^{D-2}$ so that $j \propto r^{D-3}$.  This correspondence also points out that the dilaton has a geometrical interpretation as the area of a sphere at fixed $r$. More generally, Cadoni\cite{cadoni91}showed for power law potentials $V\sim\phi^{-b}$ that black hole solutions exist providing $-1\leq b<1$. The dilaton potential for spherically symmetric gravity  in $D$ spacetime dimensions is of this form with $b=1/(D-2)$.

The energy of black hole solutions can be written in covariant form using the Killing field (\ref{killing field}). In particular, the mass observable $M$:
\be
2M = j(\phi) - |\nabla k|^2,
\label{mass observable}
\ee
is constant on-shell and a Hamiltonian analysis \cite{louis-martinez93} confirms that it corresponds to the ADM mass.

From the expression for the mass observable (\ref{mass observable}), one can readily derive the thermodynamic properties such as the temperature, surface gravity and entropy. Clearly the horizon location is a surface of constant dilaton field $\phi=\phi_h$ given by:
\be
j(\phi_h)=2M.
\ee
Variation of the above gives the analogue of the ``first law of black hole mechanics'' in this simple context:
\be
\delta M = J_{,\phi}(\phi_h)\delta \phi_h=V(\phi_h)\delta\phi_h,
\ee
where $V$ is the dilaton potential. A direct calculation of the black hole surface gravity gives:
\be
\kappa \propto V(\phi_h).
\ee
From this one can identify the value of the dilaton at the horizon as the analogue of the black hole entropy. This is to be expected from the fact that in dimensionally reduced spherically symmetric gravity $\phi = r^{D-2}$ is the area of a surface of constant radius. It can also be verified directly from the 2-d theory by using Wald's method\cite{wald93}.

The analysis above shows the beauty, simplicity and universality of the formalism: the thermodynamic properties of black holes are fundamental and completely generic\footnote{For a recent review of the thermodynamics of 2-d black holes see Grumiller and McNees \cite{grumiller07}.}. The diffeomorphism invariance allows for the possibility that generic dilaton gravity may help to provide a deeper understanding of the microscopic source of the thermodynamics, which is likely rooted in the quantum structure of the diffeomorphism group. We will now show that the subtle blend of simplicity and underlying complexity is also manifest in the quantum theory.

\section{Hamiltonian Structure and Reduced Theory}

In the following, we present a summary of the canonical quantization of generic dilaton gravity in terms of geometric variables following Louis-Martinez et. al. \cite{louis-martinez93}. We again emphasize that this class of theories has been quantized by a variety of authors. The Vienna group used the very elegant Poisson-Sigma Model approach to perform a complete classification of the classical solutions of the model coupled to a Yang-Mills field \cite{strobl96}. They were able to classify all the global solutions for the generic model and determine physical quantum states, determining the mass spectra in some cases. Interesting results have also been obtained via path integral methods (see for example the references cited in \cite{path}).

The first step is a Hamiltonian analysis, which has been well documented in the literature, so we give only essential details. The metric is first parametrized in ADM form:
\be
ds^2=e^{2\rho}\left(-\sigma^2 dt^2+\left(dx+N dt\right)^2\right).
\label{adm metric}
\ee
which leads to the action:
\be
I[\rho,\phi] = \int d^2x \left(\Pi_\rho\dot{\rho}+\Pi_\phi\dot{\phi}-\left(\sigma{\cal G} + {N}
 {\cal F}\right) -H_{B}\right),
\ee
where $H_B$ is the boundary term needed to make the variational derivative of the action well defined, the lapse $\sigma$ and shift $N$ are Lagrange multipliers that enforce the Hamiltonian and diffeomorphism constraints, respectively:
\bea
{\cal G} &:=& {\phi''\over G} -{\phi'\rho' \over G} - G \Pi_\phi\Pi_\rho -{e^{2\rho}\over2G}{ V(\phi)\over l^2}
\approx 0,\label{hamiltonian constraint}\\
{\cal F} &:=& \rho' \Pi_\rho - \Pi_\rho' + \phi' \Pi_\phi
     +\psi' \Pi_\psi\approx 0.
\label{diffeo constraint}
\eea
The presence of the two Lagrange multipliers and two first class constraints means that the number of degrees of freedom are (heuristically): 3 metric components + 1 dilaton - 2 lagrange multipliers - 2 constraints = 0. This only applies to the field theoretic degrees of freedom. It can be shown that the mass function:
\be
{\cal M} = {l\over 2G}\left(e^{-2\rho}((G\Pi_\rho)^2 -(\phi')^2)
  +{j(\phi)\over l^2}\right),
\label{mass function}
\ee
commutes with the constraints and hence is a physical observable in the Dirac sense. It is spatially constant on the constraint surface and is equal to the boundary term $H_B$ which is the ADM mass of the system.

The momentum canonically conjugate to the mass observable $M$ (\ref{mass observable}) can be written covariantly\cite{gegenberg94a} as an integral over the Killing vector field $k^\mu$, and corresponds generically to the Schwarzschild time separation at infinity as shown \cite{kuchar94} in the case of spherically symmetric 4-D gravity \footnote{An earlier derivation of the phase space observables for spherically symmetric gravity was given by Thiemann and Kastrup\cite{thiemann93} in terms of Ashtekar variables.}. The complete phase space is therefore two dimensional and can be coordinatized by the mass, $ M$ and its canonical conjugate, $P_M$. For suitably chosen boundary conditions, the fully reduced action given in terms of $M$ and $P_M$ is simply:
\be
I_{red}= \int dt \left(P_M\dot{M}-M\right).
\label{reduced action}
\ee
The resulting equations of motion imply that $M$ is time independent and that $P_M$ is equal to the time coordinate. This elegant coordinatization, while geometrically motivated, nonetheless makes it difficult to extract further information about the system without further assumptions.

\section{Semi-Classical Arguments and the Area Spectrum}

We now show that it is possible to use the laws of black hole mechanics/thermodynamics to make very intriguing general arguments about the generic semi-classical black hole mass/area spectrum in terms of the fully reduced phase space variables. Bekenstein\cite{bekenstein74} and then Bekenstein and Mukhanov\cite{bekenstein_mukhanov95} conjectured that the area of 4-D black holes is an adiabatic invariant whose semi-classical spectrum must, by the Bohr-Sommerfeld quantization condition, be equally spaced. A modern, completely general, rationale (it is not quite a proof) for this claim goes like this:
Suppose
that there is a natural frequency, $\omega$, associated with the dynamics of Schwarzschild-like black holes, which by definition have a single horizon completely parametrized by a single dimensionful variable that can without loss of generality be taken as the mass or energy, $E$.  Although at first glance the notion of oscillatory motion of event horizons may seem far fetched since one normally thinks of black holes as static with no dynamics whatsoever, we will see below that there do exist candidates for such black hole vibrational frequencies. Moreover, as argued above, the frequency $\omega(E)$ is a function of at most the energy of the black hole.

Generally, a dynamical system with an energy dependent natural frequency $\omega(E)$ has an associated adiabatic invariant that is given up to an additive constant by the indefinite integral:
\be
I=\int \frac{dE}{\omega(E)}.
\label{adiabatic invariant}
\ee
By virtue of the Bohr-Sommerfeld quantization condition, the semi-classical energy spectrum is given by :
\be
I = n h,\,\qquad n>>1.
\label{spectrum}
\ee
 Incidentally, the above argument is equivalent to the Bohr correspondence
principle which states that for large quantum numbers the classical frequency is proportional to the change in energy due to a quantum transition between adjacent states: $\Delta E = \hbar \omega(E)\Delta n$. For large $n$, $\Delta n=1$ can be treated as infinitesmal which immediately implies the differential forms of (\ref{adiabatic invariant}) and (\ref{spectrum}).

In the mid-nineties, it was noticed by two groups\cite{kastrup96,barvinsky96} that there is a natural candidate for such
 an oscillation frequency, namely:
\be
\omega(E) = \frac{k}{\hbar} T_{BH}.
\ee
This frequency corresponds to the inverse of the period in imaginary time of the Euclidean Gibbons-Perry instanton solution. An application of the semi-classical argument above\cite{kastrup96} or a direct quantization of the reduced Hamiltonian in Euclidean time\cite{barvinsky96} (\ref{reduced action}) both yield an equally spaced area/entropy spectrum:
\be
A = 2\pi (n+1/2)\hbar G.
\label{barvinsky spectrum}
\ee
Analoguous arguments were later applied to the quantization of charged \cite{das01} and rotating\cite{medved02} black holes.

An intriguing proposal for the vibrational frequency of black holes came from Hod \cite{hod98} in 1998 who argued that the frequency of the highly damped quasinormal modes of black holes are the vibrational frequencies that determine the semi-classical quantum spectrum for black holes. This proposal not only gave an equally spaced area spectrum, but the spacing that results from this choice has tantalizing consequences:  the Bekenstein-Hawking entropy takes the form $S=k\ln(3^n)$, which is consistent the statistical mechanical entropy that one might associate with an event horizon built out of $n$  elements of area, each with three allowed microscopic states.  Hod's argument went more or less unheeded until Dreyer\cite{dreyer03} showed that such a spectrum with precisely this spacing followed from LQG. Hod's argument is even more compelling in the light of the elegant calculations of Motl and Neitzke\cite{motl}. Using WKB methods, they were able to obtain analytic expressions for the highly damped QNM's of Schwarzschild black holes in higher dimensions that  were consistent with  a generalization of Hod's conjecture\cite{gk03}. This lead to a veritable cottage industry in highly damped QNM calculations for a large variety of black holes that were designed to test the universal applicability of area spectrum derived by Hod.
The results were somewhat mixed. In addition, it turned out that Dreyer's  analysis  was based on a incorrect expression for the entropy of LQG black holes\cite{ashtekar98} that has since been corrected\cite{lewandowski04}. The new expression seems on the face of it inconsistent with Hod's conjecture, but a slightly different interpretation of what one means by spherically symmetric black holes in the LQG context can be used to bring the two expressions back in line \cite{smolin06}.

The argument for the universal applicability of Hod's conjecture was revitalized by very recent paper of Maggiore\cite{maggiore07}, who used an analogy with damped harmonic oscillators to argue that the physical black hole frequency is not the real part of the QNM frequency. Instead, the physically relevant frequency is given by:

\be
\omega_0=\sqrt{\omega_R^2+\omega_I^2} ~,
\ee
where $\omega_R$ ($\omega_I$) are the real and imaginary parts, respectively. In this case, the imaginary part, not the real part, dominates the expression in the high damping limit. Moreover, $\omega_I$ appears to be more or less universal, since it is determined by the periodicity of the Euclidean instanton solution. The spectrum that results from Maggiore's reinterpretation is precisely (\ref{barvinsky spectrum}).

The unifying theme that has emerged is that all semi-classical quantization schemes appear to give an equally spaced area spectrum for Schwarzschild-like black holes in the semi-classical limit, albeit with different spacings. The apparent universality can be understood at a very basic level by noting that for Schwarschild-like black holes that are completely specified by a single dimensionful parameter (the mass/horizon radius/surface gravity), the only natural time scale is the time for light to cross the horizon, or $2GM/c^3$, whose inverse gives a frequency that is proportional to $kT/\hbar$. This fact, plus the first law of black hole thermodynamics, invariably produces an equally spaced area spectrum.
\footnote{In a somewhat different vein, Louko and Makela\cite{LM96} performed a  rigorous quantization of a reduced theory in which the radius of the throat of the Einstein-Rosen bridge provided the physical observable. Although strictly speaking there was no periodic motion in this model, they also obtained an equally spaced area spectrum, presumably for reasons that could be traced to the dimensional arguments above.} A strong hint that this argument is physically relevant (despite our uncertainty about what precisely constitutes the physical vibration frequency of a black hole) comes from the recent work \cite{lqg_spectrum} wherein the LQG spectrum is calculated.  They find a periodic structure, not unlike interference fringes, in the spectrum. This periodic structure is highly suggestive of an equally spaced area spectrum, and provides startling evidence for the relevance of the above semi-classical arguments.

\section{Dirac Quantization}

We have shown that despite the lack of local degrees of freedom, the completely reduced theory can contain some interesting  semi-classical  information about the area/entropy spectrum of black holes. However, to gain insight into the microscopic source of black hole entropy it is necessary to go deeper. By going to the completely reduced theory, one might be throwing out the baby with the bathwater. Carlip has argued\cite{carlip04} that the entropy of black holes can be understood as a consequence of boundary conditions at the horizon which effectively break diffeomorphism invariance and result in ``would-be gauge degrees of freedom'' becoming physical. A few years ago, he examined this issue in the context of generic 2-d dilaton gravity and found\cite{carlip04} that indeed the black hole boundary conditions at the horizon resulted in a modified (anomalous) constraint algebra which, when quantized, provided the right number of microstates to account for the black hole entropy.

Remarkably, Professor Kummer's last paper on the subject of 2-D dilaton gravity\cite{kummer05} shed a different and interesting new light on this type of analysis. In this paper it is shown that physical degrees of freedom on the horizon can, by imposition of horizon constraints, be converted to gauge degrees of freedom, in agreement with a conjecture by 't Hooft\cite{thooft04} and in apparent contradiction with the results of Carlip.  As argued by Bergamin {\it et al}\cite{bergamin06}, the two sets of results can in some sense be viewed as complementary descriptions of entropy in terms of inaccessible states, but clearly more work is required to understand this issue fully. It is therefore useful to explore the quantum behaviour of unreduced generic 2-D dilaton gravity.
\subsection{Exact Dirac Wave Functionals}
It turns out to be possible in the generic theory to write down candidates for exact mass eigenstates  which solve the quantized constraints\cite{gegenberg93}. These solutions were first found in the specific case of Jackiw-Teitelboim gravity by Henneaux\cite{henneaux85}.

  In terms of the ADM parametrization given in Eqs.(\ref{adm metric}), Dirac quantization in the Schrodinger representation entails a search for functionals $\Psi_{M}[\phi,\rho]$ that satisfy the quantized version of the diffeomorphism and Hamiltonian constraints.
Following Henneaux\cite{henneaux85} one first solves the two constraints classically to obtain an expression for the momenta in terms of $\rho$ and $\phi$:
\bea
\Pi_\phi &=& \frac{g[\phi,\rho]}{Q[\phi,\rho;M},\nonumber\\
\Pi_\rho &=& Q[\phi,\rho;M],
\label{classical momenta}
\eea
where
\bea
g[\phi,\rho] &=& 4\phi'' -4\phi'\rho'+2e^{2\rho} V(\phi)\nonumber\\
Q[\phi,\rho] &=& 2\sqrt{(\phi')^2+(2M - j(\phi))e^{2\rho}}\,\, .
\label{g and Q}
\eea
$M$ is a constant of integration that corresponds, as the notations suggests, to the black hole mass.
If one replaces the conjugate momenta by the standard operators:
\bea
\hat{\Pi}_\phi&=&-i\hbar\frac{\delta}{\delta\phi(x)},\nonumber\\
\hat{\Pi}_\rho&=&-i\hbar\frac{\delta}{\delta\rho(x)}.\nonumber\\
\eea
it is straightforward to integrate the quantum version of (\ref{classical momenta}). The result is:
\be
\Psi_M[\phi,\rho;M] = \exp\left(\frac{i}{\hbar}S[\phi,\rho;M]\right),
\label{wave functional}
\ee
where:
\be
S[\phi,\rho;M] = \int dx\left[Q +\phi'\ln\left(\frac{2\phi'-Q}{2\phi'+Q}\right)\right].
\ee
The wave-functional $\Psi_M$ satisfies the quantum diffeomorphism and Hamiltonian constraints with a particular (non-standard) choice of factor ordering and is an eigenstate of the quantum version of the mass function, again with a particular factor ordering.

Although the interpretation of $\Psi_M$ as an exact physical mass eigenstate has difficulties related to the choice of functional measure and self-adjointness of the relevant operators, the phase $S$ has a natural and unambiguous interpretation as the Hamilton-Jacobi function for the classical theory that derives from the classical constraints. Thus, the wave-functional (\ref{wave functional}) is at least correct to lowest order in the WKB approximation and has some interesting quantum properties: the phase is imaginary in the classically forbidden regions, $Q^2<0$ and $4\phi'^2-Q^2<0$. The latter corresponds to the region below the horizon, again forbidden along a Schwarzschild slice.
These imaginary parts were given an interesting, albeit speculative, interpretation\cite{gegenberg93} for classically forbidden configurations of $\rho$ that correspond to Schwarzschild slices with a mass parameter, $m$, different from the mass eigenvalue $M$ of the wave functional. By defining the probability amplitude for the black hole in an eigenstate of the mass function with eigenvalue $M$ to have mass $m\neq M$ as:
$P[M] \propto |\psi_M[m;M]|^2$, it was found that the relative probability of having mass $m=M$ to having no mass $m=0$ was:
\be
\frac{P[m=M]}{P[m=0]}=\exp 2\pi \frac{M^2}{m_{pl}^2}.
\ee
This expression can be interpreted as the inverse of the tunnelling probability from a state with mass $M$ to the vacuum state. Remarkably, it is proportional to the exponential of the black hole entropy (up to a factor of two), so that this interpretation is consistent with the fact that in statistical mechanics the exponential of the entropy is equal the number of accessible microstates.
\subsection{Partially Reduced Theory}
It is possible to implement a procedure that is part way between complete gauge fixing at the classical level and the Dirac quantization of the completely unreduced theory. Since one expects the resolution of many key issues in quantum gravity to reside in the Hamiltonian constraint, it makes sense to choose a partial gauge fixing which eliminates only the diffeomorphism constraint and leaves the Hamiltonian constraint to be implemented as an operator constraint via the Dirac prescription.  A few years ago Husain and Winkler\cite{hw} started a program designed to formulate the quantum dynamics of black hole formation in four dimensions. They partially fixed the gauge so as to allow slicings that were regular across the horizon. Their boundary conditions were consistent with the so-called ``flat slice'' or Painleve-Gullstrand (PG) coordinates. A similar program was initiated \cite{gks06} for the generic theory in which the analogue of PG coordinates takes the form:
\be
ds^2 = j(\phi)\left( -dt^2 + (dx + \sqrt{2GMl\over j}dt )^2\right),
\ee
The partial gauge choice was therefore $\phi'(x) =j(\phi)/\ell$, which when the diffeomorphism constraint is imposed strongly and the corresponding Lagrange multiplier, i.e. the shift function, is fixed so as to preserve the gauge fixing condition, leaves a partially reduced action of the form:
\bea
I&=&\int dxP\dot{X} - \int dx\left(-{\sigma X^2 \over 2 j(\phi)} {\cal M}'  \right)+\int dx ({\sigma X^2\over j(\phi)}{\cal M})',\nonumber\\
\label{action}
\eea
where we have done a canonical transformation from $\rho,\Pi_\rho$ to $X=e^\rho$ and its conjugate $P$, and $ M$ is again the mass function, which in this class of partially-fixed gauges is:
\be
{\cal M}:= {l\over 2G}\left(P^2 - {j(\phi)^2\over X^2}+{j(\phi)\over l^2}\right).
\ee
One can now satisfy the Hamiltonian constraint quantum mechanically by finding eigenstates of the mass function. Remarkably, the chosen partial gauge fixing results in a mass function that no longer couples different spatial points, so that the eigenvalue problem reduces to a set of decoupled quantum mechanical systems, each of which corresponds to that of a particle moving in an attractive $1/X^2$ potential. The quantization of the $1/X^2$ potential has been extensively studied in part because of the scale invariance and $SO(2,1)$ symmetry algebra that are broken at the quantum level.

The eigenstates of the mass function were found using two distinct quantization schemes, with interesting, but somewhat distinct results: Bohr, or polymer quantization\cite{gks06} for fixed mass $M$ forced a non-trivial discretization of the spatial slice: $j(\phi)$ can take on only a countable infinity of discrete values. Schr\"odinger quantization\cite{louko06}, on the other hand, yielded solutions to the quantum Hamiltonian constraint in terms of (generalized)
eigenstates of the ADM mass operator and allowed the specification of a physical inner
product in such a way as to guarantee self-adjointness of the time operator affinely
conjugate to the ADM mass. The interesting result there was that regularity of the time operator across the
horizon gave rise to a factor ordering term that
distinguished the future and past horizons, and gave rise to a quantum
correction to the black hole surface gravity.

\section{Conclusion}

Our discussion so far has dealt with vacuum 2-d dilaton gravity. Despite its underlying simplicity, it has the potential to yield significant insights into the underlying microscopic quantum theory.
Of course, in order to examine important issues such as Hawking radiation, the end-point of gravitational collapse and the quantum dynamics of black hole formation it is necessary to add matter. This is a difficult problem for arbitrary matter couplings but is tractable in the case of conformal coupling\cite{cghs}. A first step in this direction for general matter couplings in the generic theory has been taken recently \cite{hw, dgk07} by deriving the gravity-matter Hamiltonian for a massless scalar field with partial gauge fixing $\phi'=j(\phi)/\ell$. This Hamiltonian takes a rather simple and suggestive form:
\bea
H(X,P,\psi,\Pi_\psi)&=&\int dx\left(-{\sigma X^2 \over  j(\phi)} {\cal M}' +\sigma {\cal G}_M +\sigma l  \frac{XP \psi'\Pi_\psi}{ j(\phi)}  \right)+\int dx ({\sigma X^2\over j(\phi)}{\cal M})',\nonumber\\
\label{ham2}
\eea
where $\cal{G}_M$ is the matter energy density:
\be
{\cal G}_M := {1\over 2} \left( {\Pi_\psi^2\over h(\phi)} +
   h(\phi)(\psi')^2\right).
\label{cal G}
\ee
There is no space to describe this model in detail, but this Hamiltonian has some potentially useful properties. There is a clean separation between the pure gravitational sector (the first term), the matter sector (second term) and a quartic interaction (third term). This rather simple form is a direct consequence of our partial gauge fixing: the function $\phi(x)$ is no longer dynamical but a fixed function of the spatial coordinates.

One possible approach for solving the Hamiltonian constraint may be as follows: one can take as a hopefully complete basis the eigenstates of the mass function found via Bohr quantization in \cite{gks06} or via Schr\"odinger quantization in \cite{louko06}. In addition, one can use standard techniques to find a complete basis of states for the scalar field using just the matter term in the Hamiltonian constraint.
The interaction term can then be formally expressed in the corresponding direct product basis, allowing the Hamiltonian constraint to be solved using perturbative techniques. Perturbation theory will likely not be valid near the singularity but may be relevant near the horizon of macroscopic black holes. One can thus hope to address interesting questions related to Hawking radiation, including the emergence of the standard semi-classical approximation and quantum corrections to geometrical quantities such as surface gravity.
\\[5pt]

{\bf Acknowledgements:} This work was supported in part by the Natural Sciences and Engineering Research Council of Canada.


\end{document}